\documentclass[11pt]{article}
\usepackage{float}
\usepackage{caption}
\usepackage{subcaption}
\usepackage{array}
\usepackage{cool}
\usepackage{bbold}
\usepackage{eulervm}
\usepackage{hyperref}
\usepackage{tabularx}
\hypersetup{colorlinks=true,linkcolor=blue,citecolor=blue}
\let\oldmb\mathbold
\protected\def\mathbold{\oldmb}
\usepackage{amsmath}
\usepackage{graphicx}
\usepackage{color}
\begin{document}
\title{\bf Photonic band-gap transmission map of graphene in a defective
optical structure}

\author{ Dariush Jahani \thanks { e-mail address:  dariush110@gmail.com } }
 \maketitle {\it \centerline{  \emph{}}
\begin{abstract}
\emph{{ Interaction of graphene with a defect layer of a 1D defective photonic structure correspondingly could open a Dirac gap in its nonlinear energy dispersion. Also, an excitonic gap could be created upon applied strong magnetic fields at low temperatures. Now, the exact solution for the transmission of the allowed defective states under the influence of a homogeneous magnetic field requires providing a new scheme for the transfer matrix method to cover the circularly polarized propagation of the light in THz regime. In this paper, through an easy-to-follow framework for obtaining the distinct expression of the related transfer matrix method, it is observed that the hybrid defective states in the suggested optical structure undergo an unconventional transmission map as function of Dirac gap. Moreover, it is shown that mapping the transmission spectrum in the frequency space reveals peculiar results for narrow-band transport of the defective states. Finally, it is demonstrated that the transmission spectrum shows plateaus as a function of the applied magnetic field analogues to the situations observed for the conductance of a 2D electron gas system in the quantum Hall effect regime. This almost simple suggested design which can be fully controlled marks the first optical counterpart of quantum Hall effect in optics.
 }}
\end{abstract}
\vspace{0.5cm} {\it \emph{Keywords}}: \emph{Transmission spectrum; Defective photonic crystal; Gapped graphene, Quantum Hall
defect modes. Optical quantum Hall systems.}
\section{Introduction}
\emph{In quantum electronics, the quantum Hall effect (QHE) observed for low-dimensional 2D electron systems for the Hall conductance, apart from the usual quantization which is expected in the quantum physics, shows step-like transitions [1]. The key point toward observation of these step-like scheme is the existence of defect states subjecting the charge carriers of the 2D electron gas to be then strongly sensitive to the applied magnetic field [2,3]. In 2D materials, therefore, the underlying physics behind the outer shell-electronics of the system could be reflected under Quantum Hall regime. A well-know example certainly is the recent observation of the unconventional quantum Hall effect in graphene, a 2D carbonic material packed into a honeycomb lattice for which the Hall experiment proved its massless Dirac carriers for the first time [4-6]. These massless electrons in the low energy limit are responsible for the emergence of the Dirac cones in the linear electronic spectrum of graphene and might be of interest for ultra-fast electronic transport communications. Furthermore, the interaction of the electromagnetic field with the ultra-relativistic massless electrons moving with constant velocity in graphene could be of great interest in high speed optoelectronics applications [7,8]. }
\par
\emph{In quantum optics, optoelectronic properties of graphene in different situations have been subject of considerable investigations [9-12]. Among these studies, the optical conductivity of graphene in the presence of a constant magnetic field for which the intra- and inter-band transitions play a significant role in optical properties of graphene has also been evaluated [13-15]. Furthermore, some resent progresses also have been made in order to study the optical QHE systems in graphene [16,17]. Now, considering the interaction of graphene as a covering material of the defect layer of a 1D defective optical structure which might lead to an opening gap in its linear spectrum, excite motivations to seek the band-gap transmission map of the allowed circularly polarized states due to the coupling effect of the magnetic field on TE and TM polarizations [18]. Further, up to this stage, no work has reported on the emergence of the EM propagation undergoing a step-like transmission map analogous to the plateaus which is observed in the electronic QHE regime reported. Thanks to the transfer matrix method developed in in this rather fast communication based on an analytical framework for the circularly polarized defective states, for the first time in optics, the most general but explicit aspects of a step-like transmission as a function of the magnetic field in a 1D defective optical structure is addressed which could be of most interest in fast communication optics and nano-photonics }
 \par
\emph{Now, in photonic structures, with broken periodicity transport of light waves are capable of showing properties similar to a periodic atomic structure with defect states arising from the propagation of its electronic waves in a partial-symmetry broken lattice with a associated finite effective mass [19,20]. In particular, defect modes propagation passing through defective photonic crystals (DPCs) with a background of forbidden transmission could be emerged just analogous to the creation of the defect energy levels in electronic systems. Interestingly, resonances for transmission are not expected to occur for defect modes and, therefore, these narrow-band propagation modes are more capable of being controlled by additional parameters introduced in the structure. Opening a gap which could be created upon the interaction of grown graphene on a substrate acting as a defect layer (D-layer) in a 1D defective structure could play a central role in the control of the circularly polarized defect modes as an additional parameter. This is because it can modulate the pattern of LLs even for a constant magnetic field. The attention should also be drawn to this point that since the D-layer is covered by graphene the tunability of the defects modes are strongly dependent on the optical conductivity of the gapped graphene [21-23]. Moreover, simple one-dimensional geometry of the structure introduced in this work, certainly, introduces a forbidden transmission background for the emerging circularly defect modes for which as it was mentioned no resonances are expected. One expects, now, different parameters affect the height and the position of defects states in the corresponding filtered background. To simulate the behavior of these defective states a new scheme for the transform matrix method is required to accept the hybrid polarization of light due to the coupling effect of the magnetic field for both linear TE and TM polarizations of surface waves in graphene. Consequently, in this situation, transitional behavior for the transmission due to distinct pattern of of LLs in graphene could be expected to be observed in the corresponding simulations for defect states in QHE regime.} \par
\emph{The aim of this paper is then to simulate the explicit band-gap transmission map of circularly defect modes associated with a 1D DPC with a D-layer interacting with graphene since it comes to a more realistic configuration when the effect of $SiC$ acting as the substrate for graphene could be considered. These defective modes of transmission are expected to exhibit a transitional scheme which might reflect LLs quantization in QHE regime for the electronic transport in graphene with a nonlinear spectrum due to the presence of the interactions between D-layer and grown graphene. Therefore, transfer matrix method for hybrid modes should be analytically developed before one can go through the numerical approach for this problem. In this regard, in section .2, this letter starts with obtaining the exact analytical solution for the transport of coupled TE-TM waves in a 1D DPC for which the defect layer is interacting with graphene. The associated numerical simulations are then immediately followed in this section to just address the most general results and, finally, in section .3 it ends with the conclusion and closing remarks.}
\section{Theoretical framework and simulations}
  \emph{In this section, the exact solution for the transfer matrix method related to the transmission of TM and TE waves passing into a graphene sheet placed between dielectric materials in the QHE regime is obtained. Then, as it is shown schematically in Fig. 1, the introduced 1D dielectric model could be schemed like ($\epsilon_{1} \epsilon_{2} \epsilon_{1}...$) for $z < 0$ and ($\epsilon_{2} \epsilon_{2} \epsilon_{2}...$) for $z > 0$ which hosts a defect layer interacting with grown graphene sheets at $z = 0$) will be analytically analyzed. It should be noted that photonic structures hosting graphene were first proposed by O. Berman et al. [24,25]. To proceed, it should be mentioned that the emergence of the both TM and EM waves in graphene which under the influence of a magnetic field show coupling effects is possible [26,27]. Therefor, for a graphene layer between two dielectric materials, the electric field components of the light wave incident normally for $z < 0 $ could be written as:
\begin{equation}
\begin{cases}
E_{1x}=a_{1x}e^{ik_{1}z}+b_{1x}e^{-ik_{1}z}\\
E_{1y}=a_{1y}e^{ik_{1}z}+b_{1y}e^{-ik_{1}z}\\
\end{cases}
\end{equation}
 and for $z > 0$ is:
 \begin{equation}
\begin{cases}
E_{2x}=a_{2x}e^{ik_{2}z}+b_{2x}e^{-ik_{2}z}\\
E_{2y}=a_{2y}e^{ik_{2}z}+b_{2y}e^{-ik_{2}z}\\
\end{cases}
\end{equation}
It is obvious that $k =\sqrt{\epsilon} \frac{\omega}{c}$ represents the wave vector in the associated regions for normal incidence for which $\omega$ and $c$ are the angular frequency and the speed of light in the vacuum, respectively. At this point, for TM and EM waves influenced by the constant  magnetic field, the coupling matrix, $D1\rightarrow2$, must be calculated so that the following relation for coupled coefficients of circularly polarized EM waves holds.
\begin{equation}
\begin{bmatrix}
a_{1\pm}\\
b_{1\pm}
\end{bmatrix}
=D_{1\rightarrow 2}
\begin{bmatrix}
a_{2\pm}\\
b_{2\pm}
\end{bmatrix}
\end{equation}
 Here, $a_{i\pm} = a_{ix} \pm ia_{iy}$; $b_{i\pm} = b_{ix} \pm ib_{iy}$ (with $i:1 and 2$) and signs $\pm$ stands for the right- and left-handed polarizations, respectively. Then, the boundary conditions for the continuity of EM waves in the two regions influenced by the magnetic field must be employed. Hence, the continuity of the electric fields and their derivatives in the presence of an induced surface current in graphene, i.e $\textbf{J} = \sigma \textbf{E}_{2}$, with $\sigma$ representing the magneto-optical conductivity, yields:
\begin{equation}
\begin{cases}
\textbf{n}\times (\textbf{E}_{2}-\textbf{E}_{1}) \vert_{z=0}=0\\
\frac{\partial\textbf{E}_{1}}{\partial z} \vert_{z=0} -\frac{\partial\textbf{E}_{2}}{\partial z} \vert_{z=0} =i\omega \mu_{0}  \textbf{J} \vert_{z=0}
\end{cases}
\end{equation}
 The longitudinal part of the optical conductivity tensor $\sigma$ for graphene which proves to be more proper for the numerical calculations are presented in Ref [28] and read as :
\begin{equation}\begin{split}
\sigma_{0}(\omega)= \frac{e^{2}v_{f}^{2}\vert eB\vert\left(\hbar\omega+2i\Gamma \right)}{\pi i} \sum_{n=0}^{\infty}\left\lbrace\frac{\left[f_{d}(M_{n})-f_{d}(M_{n+1})\right]+ \left[f_{d}(-M_{n+1})-f_{d}(-M_{n})\right]}{\left(M_{n+1}-M_{n}\right)^{3}-\left(\hbar\omega+2i\Gamma \right)^{2}\left(M_{n+1}-M_{n}\right)}\right\rbrace \\
+ \left\lbrace\frac{\left[f_{d}(-M_{n})-f_{d}(M_{n+1})\right]+ \left[f_{d}(-M_{n+1})-f_{d}(M_{n})\right]}{\left(M_{n+1}+M_{n}\right)^{3}-\left(\hbar\omega+2i\Gamma \right)^{2}\left(M_{n+1}+M_{n}\right)}\right\rbrace
 \end{split}\end{equation}
 while the Hall conductivity term is:
 \begin{equation}\begin{split}
\sigma_{H}(\omega)=\frac{-e^{2}v_{f}^{2} eB}{\pi}\sum_{n=0}^{\infty}\left\lbrace \left[f_{d}(M_{n})-f_{d}(M_{n+1})\right]- \left[f_{d}(-M_{n+1})-f_{d}(-M_{n})\right]\right\rbrace \\
\times\left\lbrace \frac{1}{\left(M_{n+1}-M_{n}\right)^{2}-\left(\hbar\omega+2i\Gamma \right)^{2}}+ \frac{1}{\left(M_{n+1}+M_{n}\right)^{2}-\left(\hbar\omega+2i\Gamma \right)^{2}}\right\rbrace
\end{split} \end{equation}
In the above relations $M_{n} =\sqrt{\Delta^{2} +2nv_{F}^{2}|eB|h}$  shows the quantized energy levels which as it is seen could be tuned by the Dirac gap. The distribution are then controlled by $f_{d}(\varepsilon)=\frac{1}{1+exp( \frac{\varepsilon-\mu_{c}}{K_{B}T)}}$ with $K_{B}$ the Boltzmann constant. At this state, by the use of group expressions (1) and (2) along with the boundary condition (4) one can manage to arrive at the following relations:
\begin{equation}
\begin{cases}
a_{1x}+b_{1x}=a_{2x}+b_{2x}\\
a_{1y}+b_{1y}=a_{2y}+b_{2y}
\end{cases}
\end{equation}
 and
\begin{equation}
\begin{cases}
k_{1}(a_{1x}-b_{1x})-k_{2}(a_{2x}-b_{2x})=\omega\mu_{0}\sigma_{0}(a_{2x}+b_{2x})\pm\omega\mu_{0}\sigma_{H}(a_{2y}+b_{2y})\\
k_{1}(a_{1y}-b_{1y})-k_{2}(a_{2y}-b_{2y})=\omega\mu_{0}\sigma_{0}(a_{2y}+b_{2y})\pm\omega\mu_{0}\sigma_{H}(a_{2x}+b_{2x})
\end{cases}
\end{equation}
 The calculations could be proceeded by multiplying relations (7) by $±k_{1}$ and then correspondingly adding the results to the equation (8) which gives:
\begin{equation}
\begin{cases}
2k_{1}a_{1x}=k_{1}(a_{2x}+b_{2x})+k_{2}(a_{2x}-b_{2x})+\omega\mu_{0}\sigma_{0}(a_{2x}+b_{2x})\pm\omega\mu_{0}\sigma_{H}(a_{2y}+b_{2y})\\
-2k_{1}b_{1x}=-k_{1}(a_{2x}+b_{2x})+k_{2}(a_{2x}-b_{2x})+\omega\mu_{0}\sigma_{0}(a_{2x}+b_{2x})\pm\omega\mu_{0}\sigma_{H}(a_{2y}+b_{2y})
\end{cases}
\end{equation}
 and
\begin{equation}
\begin{cases}
2k_{1}a_{1y}=k_{1}(a_{2y}+b_{2y})+k_{2}(a_{2y}-b_{2y})+\omega\mu_{0}\sigma_{0}(a_{2y}+b_{2y})\pm\omega\mu_{0}\sigma_{H}(a_{2x}+b_{2x})\\
-2k_{1}b_{1y}=-k_{1}(a_{2y}+b_{2y})+k_{2}(a_{2y}-b_{2y})+\omega\mu_{0}\sigma_{0}(a_{2y}+b_{2y})\pm\omega\mu_{0}\sigma_{H}(a_{2x}+b_{2x})
\end{cases}
\end{equation}
 Finally, coupling is readily seen by the use of the complex conductivity $\sigma_{\pm} = \sigma_{0} \pm i\sigma_{H}$. Multiplying the
expressions (10) by ($\pm i$) and then adding it correspondingly to expressions (9) leads to the following equations:
\begin{equation}
\begin{cases}
2k_{1}a_{1\pm}=k_{1}(a_{2\pm}+b_{2\pm})+k_{2}(a_{2\pm}-b_{2\pm})+\omega\mu_{0}\sigma_{\mp}(a_{2\pm}+b_{2\pm})\\
-2k_{1}b_{1\pm}=-k_{1}(a_{2\pm}+b_{2\pm})+k_{2}(b_{2\pm}-b_{2\pm})+\omega\mu_{0}\sigma_{\mp}(a_{2\pm}+b_{2\pm})
\end{cases}
\end{equation}
The above relations could be also written in the following matrix form to reads:
\begin{equation}
\begin{bmatrix}
a_{1\pm}\\
b_{1\pm}
\end{bmatrix}
=\frac{1}{2}
\begin{bmatrix}
1+\frac{k_{2}}{k_{1}}+\frac{\omega\mu_{0}}{k_{1}}\sigma_{\mp}&&
1-\frac{k_{2}}{k_{1}}+\frac{\omega\mu_{0}}{k_{1}}\sigma_{\mp}\\
1-\frac{k_{2}}{k_{1}}-\frac{\omega\mu_{0}}{k_{1}}\sigma_{\mp}&&
1+\frac{k_{2}}{k_{1}}-\frac{\omega\mu_{0}}{k_{1}}\sigma_{\mp}
\end{bmatrix}
\begin{bmatrix}
a_{2\pm}\\
b_{2\pm}
\end{bmatrix}
\end{equation}
 It is by now that the hybrid transfer matrix $D1\rightarrow2$ show up itself to be:
\begin{equation}
D
=\frac{1}{2}
\begin{bmatrix}
1+\frac{k_{2}}{k_{1}}+\frac{\omega\mu_{0}}{k_{1}}\sigma_{\mp}&&
1-\frac{k_{2}}{k_{1}}+\frac{\omega\mu_{0}}{k_{1}}\sigma_{\mp}\\
1-\frac{k_{2}}{k_{1}}-\frac{\omega\mu_{0}}{k_{1}}\sigma_{\mp}&&
1+\frac{k_{2}}{k_{1}}-\frac{\omega\mu_{0}}{k_{1}}\sigma_{\mp}
\end{bmatrix}
\end{equation}
 Therefore, extending this relation to  the introduced photonic device, one can extend it to other layers ($i - 1 \rightarrow i$; $i = 2, 3, 4, ...$) noticing that generally complex conductivity $\sigma_{±}$ can be neglected in the absence of the graphene layer.}
 \par
 \emph{Here, a multiplication of transmission matrices across different interfaces is needed for obtaining the reflection and transmission of EM waves through 1D QHE DPC. this propagation matrix (P) with N the number of layers which can join the fields at $z + dz$ to the fields at $z$ position is as follows.
  \begin{equation}
P(\Delta z)=
\begin{bmatrix}
e^{-ik_{z}\Delta z}&&0\\
0&&e^{ik_{z}\Delta z}
\end{bmatrix}
\end{equation}
 Finally, the transmission spectrum for circularly defective propagations in the introduced 1D QHE
DPC is obtained in the following scheme:
 \begin{equation}
M=D_{1\rightarrow 2}P(d_{1,2})D_{2\rightarrow 3}P(d_{2,3})....D_{N-1\rightarrow N}P(d_{N,N+1})
\end{equation}
 by which, the transmission (T) and reflection (R) is given as:
\begin{equation}
R= \vert\frac{M_{21}}{M_{11}}\vert^{2}\; ;\qquad T=\vert\frac{1}{M_{11}}\vert^{2}
\end{equation}
 which provides a proper tool to theoretically investigate the transmission spectrum of the suggested device inside through the new scheme for transfer matrix method for circularly polarized defective modes. . After reading this manuscript, I found I cannot recommend it to be published in scientific report, although the idea is interesting, in that
}
  \par
 \emph{Now, the above analytical discussion provides a proper tool for simulating the behavior of hybrid defect modes for a wide range of the Dirac opening in graphene's spectrum. For this aim, the structural parameters which is more important to the simulation results are considered. The dependence of transmission on the intensity of magnetic fields, chemical potentials and temperature both for circularly right-handed and left-handed polarizations will be also evaluated. The values of the opening gap taken into consideration in the corresponding simulations would be up to $\Delta= 300 \ meV$. In this situation, mapping the defective modes results in a defective spectrums, which in this paper, are referred to as gap-transmission spectrum. Moreover, the gap-transmissions under the effect of increasing the magnetic field in QHE regime would be illustrated. Also, as a matter of more illustration the effect of gate bias on the gap-transmission spectrum will be also considered. Although, under QHE situation, high magnetic fields at low temperatures are of much interest, the role of increasing the temperature is also addressed in the numerical simulation ahead. Therefore, the value of the temperature and the chemical potential which is generally chosen to be $T = 10\ K$ and $\mu= 0.2 \ eV$, respectively, will be changed in some cases to cover some more special results. For this purpose, the room temperature behavior ($T= 300 \ K $ ) of defective modes in the mentioned gap interval considered in this work, i.e. up to $\Delta= 300 \ meV$ is simulated.
 }
  \par
 \emph{Mapping the gap-transmission spectrum for the right-handed EM defective modes is illustrated in Fig. 2. It shows the band-gap transmission modes at $B = 10$ and $B = 15$ Tesla for different values of the Dirac gap opening ranging from $0$ to $300 \ meV$. It is seen in this figure that the transmission area grows with increasing the magnetic field in some frequency ranges. This effect could be more illustrated by shifting to the frequency space where the position of the defect modes within a narrow $1\ meV$ energy interval, as shown in Fig. 2b, moves toward the lower frequencies upon higher transmissions. Therefore, stronger magnetic fields lead to higher transmissions for the defective modes in some upper parts of the band-gap transmission map. This condenses the modes to the lower positions in the frequency domain. Oscillation in transmission in this case, as it is clear, for further increasing of the gap values is observed (see Fig.2a). Correspondingly, in the frequency space, these high transmitted modes stand out in a narrow lower regions in the frequency space and oscillate upon wider opening of Dirac gaps. Now, lets see how the chemical potential affect these band-gap transmission maps. Note that it is dependent on the doping level or the gate voltage which is capable of being controlled during the growth which itself might introduce a Dirac gap. This effect is simulated in Fig. 3a for an applied magnetic field $B = 10$ Tesla. It is obvious that the transmission map shows overall reduction through the gap interval. Observing this behavior in the frequency domain in Fig. 3b reveals that increasing the gate bias leads to the filtering of low position modes at $T = 10$ $K$. This low temperature behavior of the transmission is expected to be modified as the temperature grows. The reason behind this is the thermal excitations which could perturb the well separated LLs created at relatively high field $B = 10$ Tesla. The result shown in Fig. 4 for $T = 30 \ K$, $T = 150 \ K$ and $T = 300\  K$ proves that the transmission of right-handed modes shrinks in a smoothly manner as the temperature increases. Hence, in contrast to low temperature situation, the transmission for opening Dirac gaps at the room temperature shown in Fig. 4a introduces a more smoother map for defective EM propagations. In the frequency space, however, this corresponds to spreading out the hybrid defect modes in the position as it is clear in Fig. 4b. It is by now that it might be simply understood that the separation distance between LLs plays a central role in the low temperature limit for observing the transitions in the transmission map. This is because of the EM waves responding correspondingly to the different map of separation for LLs. However, at a fixed magnetic field, left-handed defective modes show different responses to a certain separation distribution of LLs. This also holds true for the distribution of the Dirac gap. Therefore, upon a constant value for the magnetic bias left-handed defective modes could map out different gap-transmission behavior for the considered 1D DPC.
 }
  \emph{To prove different transmission spectrums for left-handed modes, the result of the numerical simulations for magnetic fields $B = 10$ and $B = 15$ Tesla are shown in Fig. 5. By comparison with the results for right-handed modes, Fig. 5a shows that there is a strong drop in the transmission within the gap interval $0$ to $200 \ meV$. However, the oscillatory parts of the transmission spectrum map out almost same values for defective modes. Correspondingly, the frequency behavior of these modes depicted in Fig. 5b reveals that the low transmission parts upon red-shifting get more broadened in the frequency through the left-handed transport. Additionally, in Fig. 6a the effect of the increasing the gate voltage on the transmission of the left-handed defect modes appearing in the proposed 1D DPC is illustrated. As it is seen for the the values $\mu = 0.1$, $µ = 0.3$, and $\mu = 0.5$ $eV$ in a fixed bias $B = 10$ Tesla, the transmission maps distinct modulations. Here, the oscillations in transmission would be more diminished upon increasing the chemical potential. Furthermore, a decrease in the transmission is seen for smaller values of the Dirac gap. Now, Fig. 6b again shows that the emerging modes at lower frequency domain reveal lower transmissions. Next, as a matter of comparison the temperature behavior of the left-handed defect modes is also examined. Not that, for the ease of the illustration the value of the magnetic field has been chosen to be $B = 20$ Tesla in Fig. 7. It is observed that for $T = 30$, $T = 150$ and $T = 300 \ K$ one can observe readily from the Fig. 7a that there is a smooth shift toward lower transmissions in this case. The spectrum in the frequency space as it is clear in Fig. 7b shows broadening at higher temperature for left-handed defect modes.
  }
  \par
 \emph{Up to this stage, the gap-transmission spectrum for hybrid defect modes in a 1D DPC was investigated for some special values of the magnetic field. Now, by considering more general values for the applied magnetic field one could prove that the transmission of defect modes, interestingly,
maps out a step-like feature which is analogues to the quantum Hall plateaus observed for the Hall conductance in 2D electron systems. Generally, for magnetic fields changing from $B = 1$ to $B = 20 T$ the transmissions for Dirac band-gap openings $\Delta = \ 30 meV$, $ \Delta= \ 90 meV$ and $\Delta = 110\  meV$, as shown in Fig.8a, undergo step-like transitions. Note that in electronic the the transmission and conductance are related by the Büttiker formula and therefore this result is of most important in optical quantum hall systems and could open new routes toward both defect modes transmission in almost exhausted area of defective photonics research and also new optical systems which works in infrared domain. Further, the transitions could also be depicted in the frequencies as shown in Fig. 8b. It is clear that the transmissions modes which correspond to each step in Fig. 8a are almost well separated in position and grow upon increasing the magnetic bias. Therefore, these quantum Hall defect modes which from now on are referred to as QHDs reveals perceptible changes upon responding to different values of Dirac gap. It should be stressed that here the right-handed QHDs modes was considered and more analyzing of QHDs, under different situations such as blue and red shifting, is the subject of another project.
}
  \section{General remarks and Conclusions}\label{section.four}
 \emph{Based on an analytical scheme developed in this paper for the transfer matrixes related to propagation of the allowed circularly polarized EM waves emerging in a 1D defective photonic device, the band-gap transmission spectrum of graphene in QHE regime for some general situations of interest was theoretically investigated. The simulations showed that the effect of nonlinearity in the graphene's dispersion relation on both right-and left-handed defect modes could be illustrated peculiarly through a band-gap transmission map. It was then observed that the transmission spectrum reveals some transitions through the gap domain and also shows an oscillatory behavior for higher transmission modes. Further investigations also revealed that the transmission map of circularly defect modes is strongly sensitive to the gate voltage under the applied magnetic field. Interestingly, simulation also revealed almost step-like transmissions for QHDs in the magnetic domain which upon increasing the magnetic field showed increasing in the magnitude. This behavior of QHDs which might be considered as an analogues to the situation commonly observed for QHE in 2D electron gas systems, could open a new route toward further investigations on high-sensitive optoelectronics communications. The band-gap transmission map of QHDs in the proposed 1D DPC might serve an strong tool for measuring and analyzing the opening gap in graphene's dispersion and other materials which is used as the covering
materials in optical defective structures. The detail of the corresponding simulation results is found in section 3, however, in what follows more general concepts and conclusion remarks are briefly addressed. }
\par
\emph{For right-handed polarized defect modes it was shown that a sawtooth feature for the transmission in the considered range of opening gaps is observed. Moreover, it was seen that the transmission gradually increased upon applying higher intensities for the magnetic field. This growing of the band-gap transmission map for defect modes was observed to be almost diminished upon a relatively high transmission step with oscillatory behavior. Correspondingly, in the frequency space, these high transmission modes associated to the oscillatory step were observed to be red-shifted and localized in the relatively lower frequencies. However, on the other hand, higher frequency domain showed extended low transmission defect modes. As was discussed, recent researches have revealed that creation of an excitonic gap under strong magnetic field intensity and low temperature is possible in graphene. Therefore, band-gap transmission spectrum of the narrow-band circularly defect modes could be a useful tool to detect any excitonic gap which could be created in graphene upon applying strong magnetic fields.  }
\par
\emph{On the other hand, left-handed polarized defect modes revealed different transmission responses to the applied magnetic bias comparing to the right-handed's spectrums. It was demonstrated that gap transmission spectrum for left-handed defect modes exhibit a low transmission sawtooth feature which could spread in the position in the frequency domain. It seems that more localization of the modes in the position indicates higher transmission for them. Interestingly, in the frequency space circularly left-handed polarization for defect modes shows almost a mirrored feature in position relative to the allowed right-handed propagation. To be more specific, contrary to the right-handed defective propagation which was seen to extend in higher part of the photon frequency and localize with higher transmission in relatively lower parts, the oscillatory step associated with left-handed modes show up in higher positions of the frequency domain relative to the extended modes. This different behavior of two polarization modes is the direct result of the constant direction of the magnetic field which points to a special orientation in the space. In this situation different response of the defect modes could either magnify or reduce the effect of the magnetic field introducing also some directional considerations. It should be noted that apart from the case of excitonic gap, generally, the magnetic field does not affect the Dirac gap, however, it could change the gap between the LLs. Therefore, changing the direction of the field has no effect on separation distance of the energy levels. Hence, it is the different responses of the right and left-handed modes to the LLs pattern that results in these broad properties for them.}
  \par
\emph{Then, the effect of increasing the temperature was considered as it can disturb LLs pattern and the band-gap transmission map could be changed accordingly. Moreover, high temperatures at higher chemical potentials could also affect the excitonic gap [29]. In fact, similar to the central role that the temperature plays in electronic QHE systems, the effect of increasing thermal excitation on the defective modes proved significant modulations for the transmission spectrum of introduced 1D defective photonic structure in this work. It was shown that the leaps in the transmission for both right and left-handed defective states gradually diminish at higher temperatures. The delocalization and, therefore, reducing the density of the modes in the frequency zone without very significant change in the transmission was another important result to be noted in this case. Further, oscillations of the circularly defect states upon the last steps showed to be diminished by increasing the temperature. Note that this thermal sensitivity of the transmission modes of the device might be of most interest when it comes to real world applications of optical thermal detectors. Moreover, the effect of the chemical potential for some special values of the magnetic field was examined. It was shown that increasing the chemical potential which could be controlled by doping on the transmission spectrum shrinks the overall transmissions of defect modes.}
\par
\emph{Finally, the magnetic control of emerging QHDs in THz region for right handed modes was discussed. The results for allowed transmission modes as a function of the magnetic field revealed some transmission plateaus similar to the situation seen for the conductance of 2D electronic gas systems in QHE regime. Increasing the field intensity in this case provided higher transmission for QHDs with larger steps. However, increasing the opening gap led to lower transmissions for QHDs in the considered domain of field intensities. Emerging of these QHDs which is for the first time in optics could simulate motivations for further investigations of quantum optical systems operating in different frequency regimes. Note that the calculated results provide more theoretical explanation and simulations format has been chosen to provide appropriate toolbar visualizing the properties of defect modes within a narrow frequency zone. It should be noted that, the existence of photonic Hall plateaus in photonics could offer new opportunities for studying photonic Hall systems since they could be easily modified into any favourable geometry alongside with using a much more wide range of materials and, therefore, possibility of being more controlled than electronic Hall systems. }
\par
\emph{To close, the band-gap transmission spectrum of grapehene related to the EM propagation states under QHE regime in a 1D defective optical system was discussed. Then, based on an exact analytical approach, numerical simulations were performed to prove emergence of circularly polarized defect modes showing special modulation in their transmission map in THz zone upon applied magnetic field. The transmission spectrum also showed to be strongly sensitive to the increasing of the temperature and chemical potential. Surprisingly, the transmission map as a function of the magnetic field in the presence of opening gaps revealed QHDs which mimic the electronic transport of 2D electron gas systems in QHE regime. This results could be simply put to the test by the experiment.}
\section{Additional information}{\emph{\textbf{Competing Interest}: The author declare no competing interests.}}

\begin{figure}
 \begin{center}
\includegraphics[width=10cm]{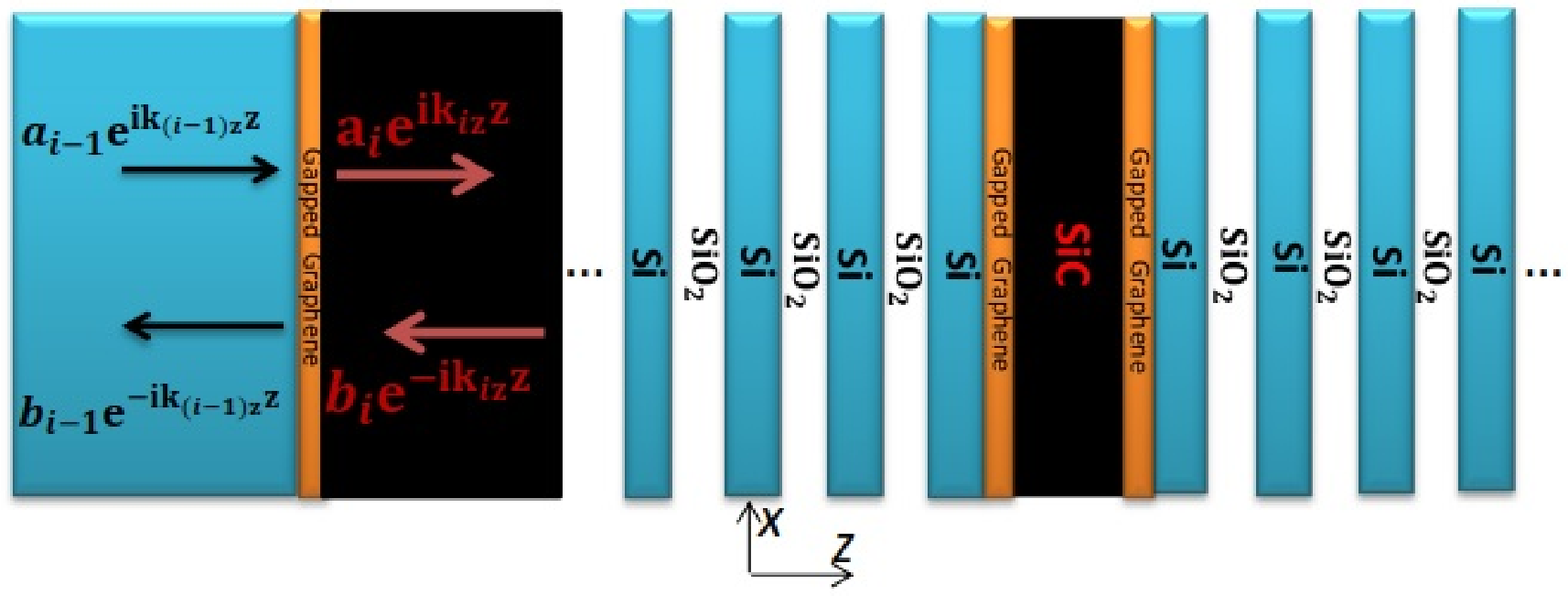}
\caption{Left: the electric field components of EM propagations in two adjacent layers for which graphene is placed between them. Right: representation of the suggested 1D defective structure with dielectric materials $Si$ and $SiO_{2}$ placed at both sides of a graphene-covered $SiC$ layer.}
\end{center}
\end{figure}
\begin{figure}	
	\centering
	\begin{subfigure}[b]{10cm}
		\centering
		\includegraphics[width=10cm]{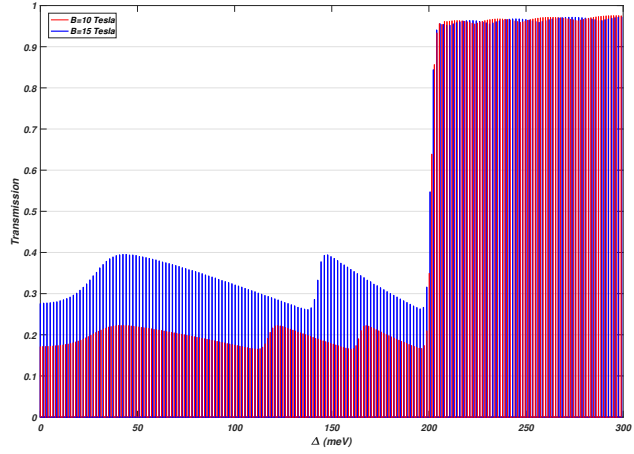}
		\caption{}\label{figur2b}		
	\end{subfigure}
	\quad
	\begin{subfigure}[b]{10cm}
		\centering
		\includegraphics[width=10cm]{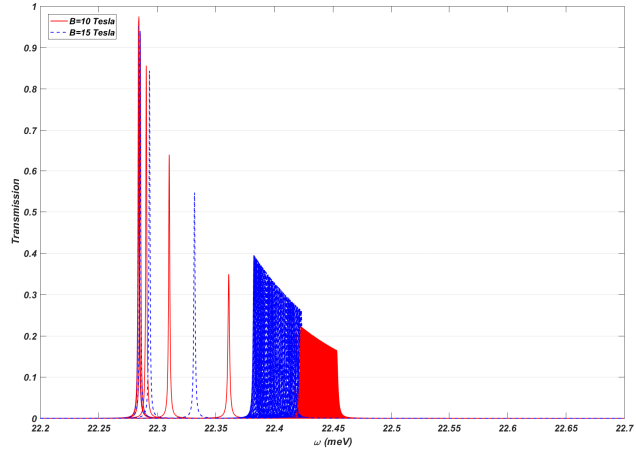}
		\caption{}\label{figur2a}
	\end{subfigure}
	\caption{Simulations results for the photonic band-gap transmission maps of right-handed modes under increasing the magnetic bias at low temperature $T = 10\  K$ for which the chemical potential reads $\mu = 0.2 \ eV$. (a) As it is clear increasing the magnetic field results in increasing the transmission before the last steps. The sawtooth-like feature is observed in this case with the peaks for the transmission near $0.4$ for $B=15\ $ Tesla and above $0.2$ for $B=10$ Tesla. (b) The photonic transmission map in the narrow frequency domain of around $1\ meV$ shows that the defective states are red-shifted toward higher transmissions upon increasing the graphene's band-gap opening. it is obvious that increasing the magnetic field also makes the spectrum undergo a narrower scheme in the frequency. }\label{fig:1}
\end{figure}
\begin{figure}	
	\centering
	\begin{subfigure}[b]{10cm}
		\centering
		\includegraphics[width=10cm]{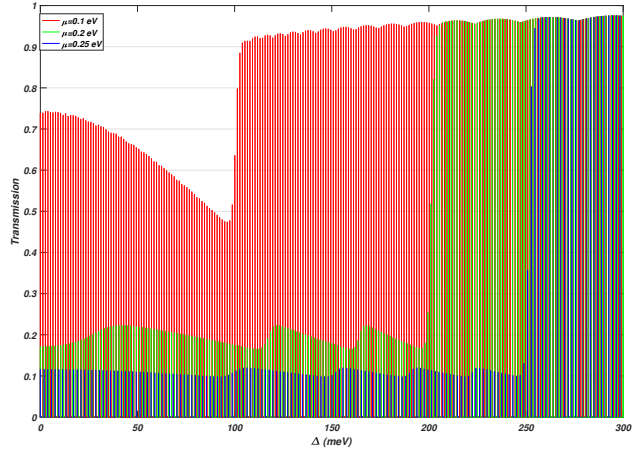}
		\caption{}\label{figur2b}		
	\end{subfigure}
	\quad
	\begin{subfigure}[b]{10cm}
		\centering
		\includegraphics[width=10cm]{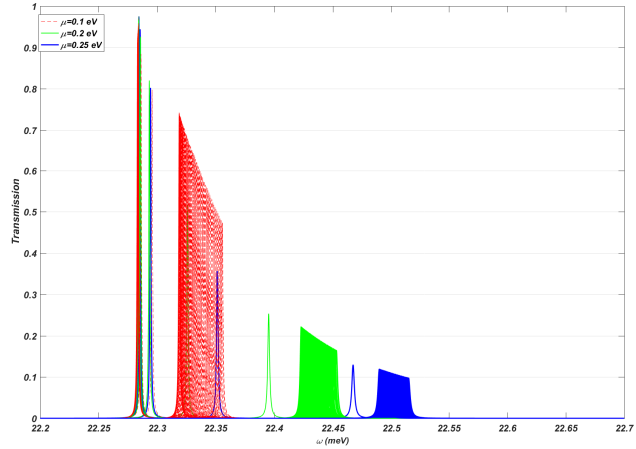}
		\caption{}\label{figur2a}
	\end{subfigure}
	\caption{ The role of chemical potential on the gap-transmission spectrum of right-handed modes for the temperature $10 \ K $ and $B = 10\  T$. (a) Further increasing of the gate voltage is shown to reduce the transmission in the sawtooth-like area. Again the peaks of the transmissions of the defect states in this area map almost similar values. Note that increasing the chemical potential increases the sawtooth-like feature of the transmission map. (b) Transmission spectrum in the frequency zone reveals that unlike the magnetic field which red-shifts the modes, the role of the chemical potential is blue-shifting the right handed modes. }\label{fig:1}
\end{figure}
\begin{figure}	
	\centering
	\begin{subfigure}[b]{10cm}
		\centering
		\includegraphics[width=10cm]{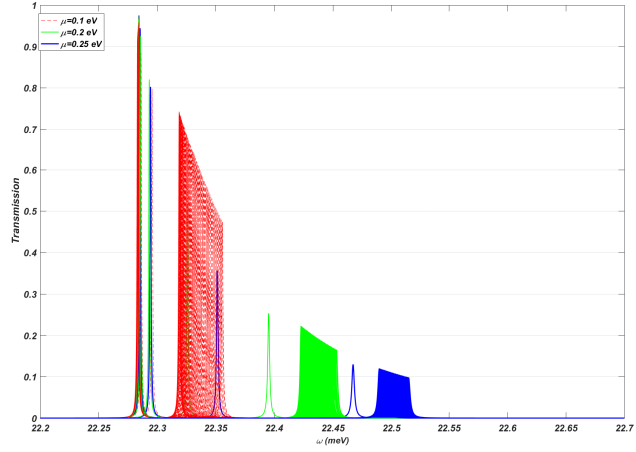}
		\caption{}\label{figur2b}		
	\end{subfigure}
	\quad
	\begin{subfigure}[b]{10cm}
		\centering
		\includegraphics[width=10cm]{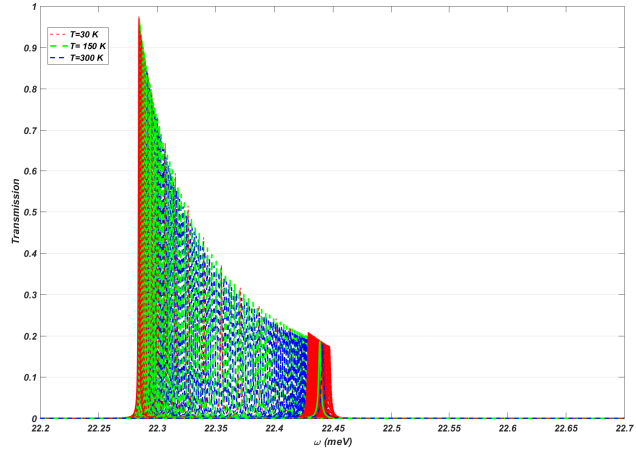}
		\caption{}\label{figur2a}
	\end{subfigure}
	\caption{The effect of increasing the temperature on right-handed propagation. (a) Band-gap transmission spectrum for temperatures $T = 30$ (red ), $T = 150$ (green) and $T = 30\  K$ (blue) under the influence of a magnetic bias $B = 10\ T$ shows that increasing the temperature dimmish the oscillatory behavior of the modes in the last transmission step as well as the sawtooth-like scheme. It also leads to the overall decreasing of the transmission specially for higher band-gap openings. (b) The transitional behavior of defect modes is also diminished and show broadening in the frequency space upon increasing the temperature.}\label{fig:1}
\end{figure}
\begin{figure}	
	\centering
	\begin{subfigure}[b]{10cm}
		\centering
		\includegraphics[width=10cm]{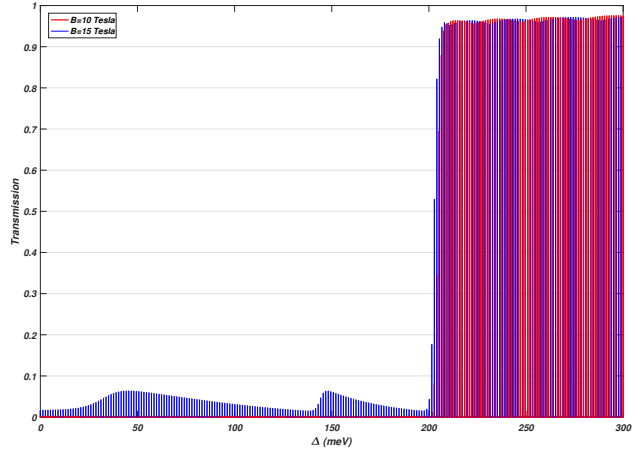}
		\caption{}\label{figur2b}		
	\end{subfigure}
	\quad
	\begin{subfigure}[b]{10cm}
		\centering
		\includegraphics[width=10cm]{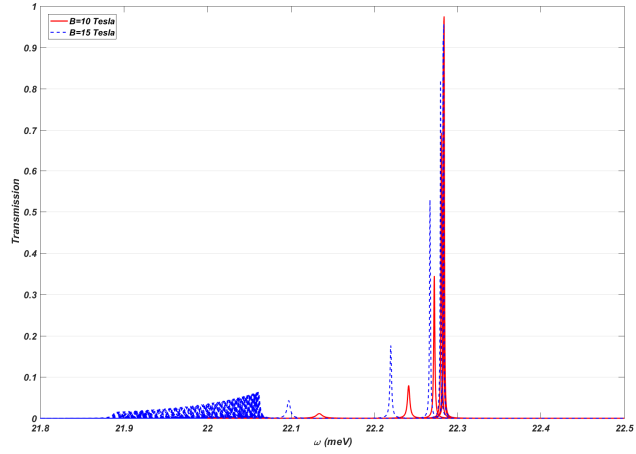}
		\caption{}\label{figur2a}
	\end{subfigure}
	\caption{Illustration of the left-handed photonic band-gap transmission map for the applied magnetic fields, $B = 10$ and $B = 15 \ T$. (a) At temperature $T = 10\ K$ and $\mu = 0.2 eV$, as it is clear, increasing the magnetic field results in overall increasing of the transmission. (b) Mapping the circularly left-handed modes in the frequency domain shows a mirrored feature in position relative to the right-handed defective propagations. In this case the modes undergo blue-shifting in the frequency domain upon stronger interactions between graphene and the SiC layer. }\label{fig:1}
\end{figure}
\begin{figure}	
	\centering
	\begin{subfigure}[b]{10cm}
		\centering
		\includegraphics[width=10cm]{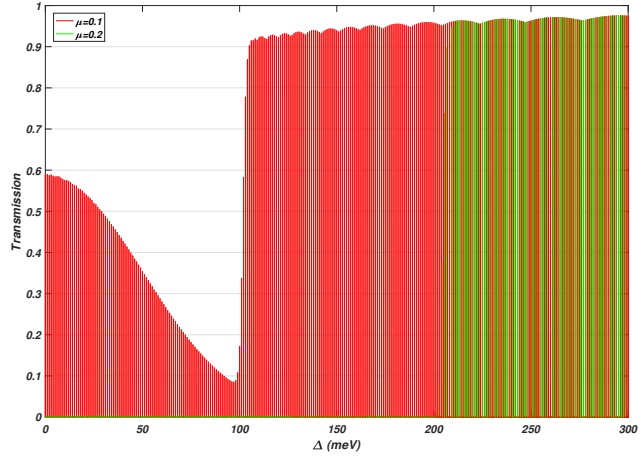}
		\caption{}\label{figur2b}		
	\end{subfigure}
	\quad
	\begin{subfigure}[b]{10cm}
		\centering
		\includegraphics[width=10cm]{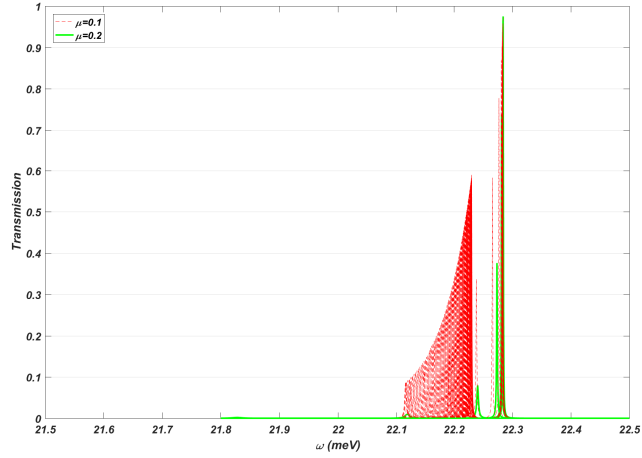}
		\caption{}\label{figur2a}
	\end{subfigure}
	\caption{The effect of the chemical potential on the transmission spectrum of left-handed modes. (a) At the low temperature $T = 10\  K$ and magnetic bias $B = 10\  T $ simulations reveal that the emerging modes for $\Delta < 200 \ meV$ are strongly filtered in the case of $\mu=0.2\ eV$. Therefore, further opening gap in graphene's spectrum makes the device more transparent for the higher doping. (b) The position-dependent behavior of the defect modes is illustrated. }\label{fig:1}
\end{figure}
\begin{figure}	
	\centering
	\begin{subfigure}[b]{10cm}
		\centering
		\includegraphics[width=10cm]{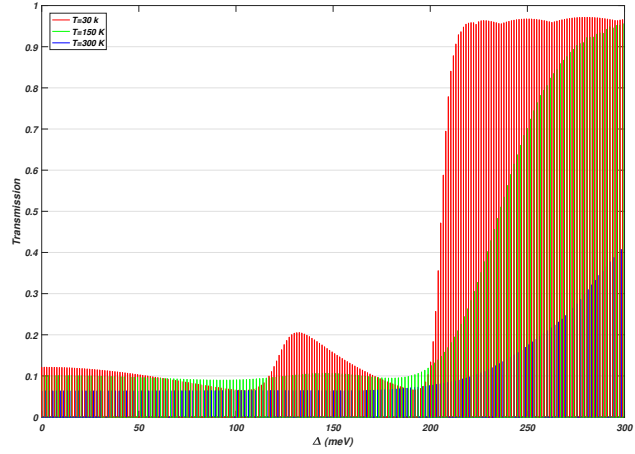}
		\caption{}\label{figur2b}		
	\end{subfigure}
	\quad
	\begin{subfigure}[b]{10cm}
		\centering
		\includegraphics[width=10cm]{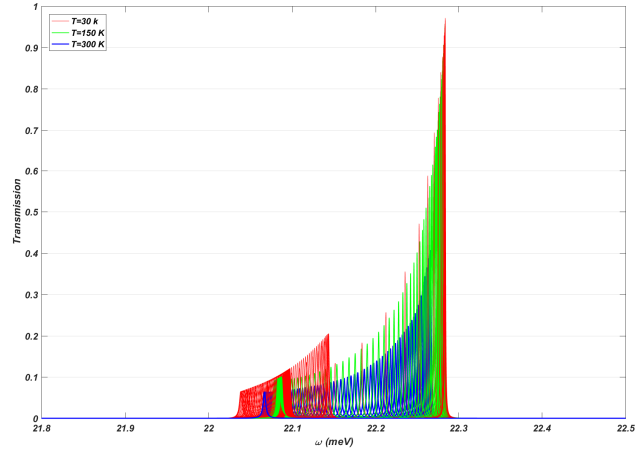}
		\caption{}\label{figur2a}
	\end{subfigure}
	\caption{Allowed temperature-dependent transmission modes for the left-handed polarizations. (a) Band-gap transmission map for temperatures $T = 30$ (red ), $T = 150$ (green) and $T = 300\  K$ (blue) under the influence of a magnetic bias $B = 20\ T$. It is obvious that circularly defect modes, generally, show lower transmission at higher temperatures. (b) The transitional behavior of defect modes is also diminished and shows broadening in the frequency space. It is interesting that increasing the temperature leads to blue-shifting of the broadened spectrum.}\label{fig:1}
\end{figure}
\begin{figure}	
	\centering
	\begin{subfigure}[b]{10cm}
		\centering
		\includegraphics[width=10cm]{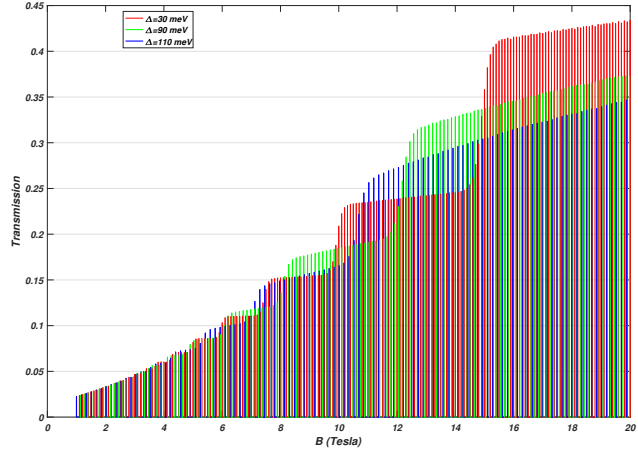}
		\caption{}\label{figur2b}		
	\end{subfigure}
	\quad
	\begin{subfigure}[b]{10cm}
		\centering
		\includegraphics[width=10cm]{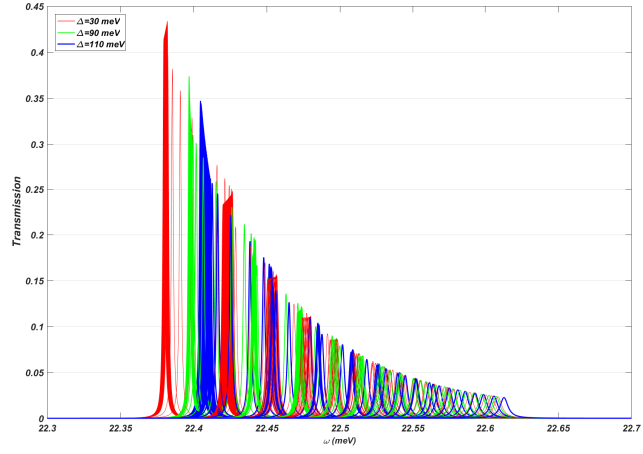}
		\caption{}\label{figur2a}
	\end{subfigure}
	\caption{Transmission spectrum for right-handed QHDs as a function of the magnetic field ranging from $B = 1 \ T$ to $B = 20\ T$. (a) Transmission plateaus for QHDs are observed which shows higher transmission steps upon increasing of the applied magnetic bias resembling the electronic plateaus seen for the electron gas systems in QHE situation. (b) The behavior of QHDs are also depicted in the frequency space. It is clear that the stronger magnetic fields red-shift the plateaus. }\label{fig:1}
\end{figure}
\end{document}